\documentstyle[prl,aps,epsfig,floats,twocolumn]{revtex}
\begin{document}
\draft
\twocolumn[\hsize\textwidth\columnwidth\hsize\csname
@twocolumnfalse\endcsname

\title{Two alternatives of spontaneous chiral symmetry breaking in 
QCD} 

\author{Jan Stern}
\address{Division de Physique Th\'eorique, Institut de Physique Nucl\'eaire,
Universit\'e\ Paris-Sud, 91406 Orsay Cedex, France\\
(IPNO/TH 97-41)}

\maketitle
\begin{abstract}

Considering QCD in an Euclidean box, the mechanism of
spontaneous breaking of chiral symmetry (SB$\chi$S) is analyzed in terms of
average properties of lowest eigenstates of the Dirac operator. A formal
analogy between the pion decay constant and conductivity in disordered systems
is established. It follows that SB$\chi$S results from a subtle balance
between the density of Euclidean quark states and their mobility. SB$\chi$S
can be realized either with $<\bar q q > =0$ , provided the low density of
states is compensated by a high mobility, or with a non-vanishing condensate,
provided the mobility is suppressed. It is conjectured that the first case
corresponds to extended whereas the latter case to (weakly) localized
quark states. 

\end{abstract}

{PACS numbers: 12.38Aw,Ly ; 11.30 Rd, Qc}

] 
\narrowtext

The subject of this note is the mechanism of flavour symmetry breaking in
confining vector-like gauge theories. We refer to QCD, but we specify neither the
gauge group (colour) nor the representation to which belong Dirac fermions
$q(x)$. This representation is assumed to be complex and there are N identical
fermion species (flavours) $q_i(x)$ with a common mass $m$. Under these
conditions the vector symmetry $U_V(N)$ remains unbroken \cite{VW84} and for
$m\to 0$ one recovers the chiral symmetry $SU_L(N)\times SU_R(N)\times U_V(1)$.
We assume the latter to be spontaneously broken. In QCD with at least three
flavours, SB$\chi$S follows from the absence of coloured physical states
 \cite{t'H80}: One proves that anomalous
Ward identities imply \cite{Col82} \cite{VW84}the existence of massless
Goldstone bosons (pions)
coupled to the conserved axial currents,
$<0|A^i_\mu|\pi^j,p>=i\delta^{ij} F_0 p_\mu$. The standard wisdom however 
goes well beyond this established fact, assuming, in addition,  that
SB$\chi$S is triggered by a formation of a large condensate $<\bar q q>$
\cite{GOR68}. No theoretical proof of this assumption is available and the
first clear experimental test of the existence of a non-negligible $\bar q
q$ condensate in the vacuum is still awaited \cite{Stern91}. Some
theoretical evidence does exist of $\bar q q$ condensation on a lattice, but
its interpretation is complicated by a triple extrapolation procedure, by
the quenching and, perhaps, by the problem of fermion doublers, which
obscures the description of SB$\chi$S within the lattice regularization.

In the following, the possible interplay between SB$\chi$S and $\bar q q$
condensation is reconsidered from the onset, exploiting the analogy between
vector-like gauge theories in continuum Euclidean space-time and disordered
systems. It will be shown that the formation of a $\bar q q$ condensate is not
the only possible mechanism of SB$\chi$S : The asymmetric vacuum and massless
Goldstone bosons coupled to the axial currents can occur even if $<\bar q
q>=0$. Nature provides examples of a similar situation: There is no symmetry
reason forcing the spontaneous magnetization of an anti-ferromagnet to vanish,
and, yet, it does vanish as a consequence of a particular magnetic order in the
ground state. Different order parameters are then necessary to describe the
system. Similarly, a complete description of SB$\chi$S in QCD may require 
order parameters other than $<\bar q q>$. We shall concentrate on the
two-point left-right correlation function
\begin{equation} \label{eq_lr}
\Delta_{\mu\nu}(q)\delta^{ij}=i \int dx e^{iqx}<\Omega|TL^i_\mu (x)R^j_\nu
(0)|\Omega>,
\end{equation}
where$L=\frac {1}{2}(V-A)$ and $R=\frac {1}{2}(V+A)$ are the Noether currents
generating left and right chiral transformations respectively. ($|\Omega>$
denotes the vacuum of the massive theory and $|0>$ stands for its
limit $m\to 0$.) In the chiral limit, the correlator (\ref{eq_lr}) must
vanish,  unless the vacuum $|0>$ is asymmetric: $\Delta_{\mu\nu}$ 
can be expressed as a
commutator of the axial charge and  represents a (non-local) order parameter.
On the other hand, if there is a Goldstone boson coupled to the axial current,
it necessarily shows up as a pole in the correlator (\ref{eq_lr}). In
particular, taking in Eq. (\ref{eq_lr}) $q_\mu =0$ and going to the chiral
limit, one obtains
\begin{equation} \label{eq_F0}
\lim_{m\to 0} \Delta_{\mu\nu} (0) = - \frac {1}{4} \eta_{\mu\nu} F^2_0 .
\end{equation}
Consequently, the expression (\ref{eq_F0}) 
represents
an order parameter of a particular type: Its nonvanishing is not only a
sufficient but also a necessary condition of SB$\chi$S.

From now on, the theory will be considered in an Euclidean box              
$L\times L\times L
\times L$ with (anti)periodic boundary conditions (up to a gauge),
and the integration over fermions will be performed first. This leads to a
quantum mechanical problem of a single quark in a random gluonic background
$G^a _\mu (x)$, which is defined by the hermitean Hamiltonian
\begin{equation} \label{eq_ham}
H = \gamma_\mu (\partial _\mu + i G^a _\mu t^a ).
\end{equation}
The result of the integration over quarks can be expressed in terms of
eigenvalues $\lambda_n$ and orthonormal eigenvectors $\phi_n(x)$ of the
Dirac Hamiltonian H. The spectrum is symmetric around the origin
: $\gamma_5 \phi_n =
\phi_{-n}, \lambda_{-n}=-\lambda_n$.  The resulting expression should then
be averaged over all gluon configurations:
\begin{equation} \label{eq_ave}
<<X[G]>>=\int d[G]exp(-S_{YM}[G])\prod _{\lambda_n >0}(m^2+\lambda_n^2)^N X[G] .
\end{equation}
The fact that this integral involves a positive probability measure suggests
a possible analogy with disordered systems. Following this way, the $\bar q q$
condensate can be written as
\begin{equation} \label{eq_cond}
<\bar q q> = -\lim_{m\to 0} \lim_{L\to \infty}\frac {1}{L^4}<<\sum_{n}\frac {m}
{m^2+\lambda^2 _n}>>,
\end{equation}
whereas the order parameter (\ref{eq_F0}) becomes
\begin{equation} \label{eq_dec}
F^2 _0=\lim_{m\to 0}\lim_{L\to \infty}\frac {1}{L^4}<<\sum_{kn}\frac {m}
{m^2+\lambda^2 _k} \frac {m}{m^2+\lambda^2 _n} J_{kn}>>,
\end{equation}
where
\begin{equation} \label{eq_JKN}
J_{kn}=\frac {1}{4} \sum_{\mu}|\int dx \phi^\dagger _k (x)\gamma_\mu \phi_n
(x)|^2.
\end{equation}
One observes that both order parameters (\ref{eq_cond}) and (\ref{eq_dec})
are merely sensitive to the infrared end of the Dirac spectrum,
$|\lambda_n|<\epsilon$. In particular, the possible ultraviolet divergences
in the sum over eigenstates of $H$ become irrelevant in the chiral limit.

A possible interpretation of
Eq.(\ref{eq_dec}) emerges,
if one considers the Dirac Hamiltonian (\ref{eq_ham}) as a generator of
evolution in a fictitious time $t$
added to the four Euclidean space coordinates
$x_\mu$. In the corresponding 4+1 dimensional space-time one can switch on
a homogeneous colour singlet electric field ${\cal E}_\mu \cos(\omega t)$,
adding
to $H$ a time-dependent perturbation
$\delta H = i\gamma_\mu {\cal E}_\mu \frac {\sin \omega t}{\omega}$,
where $\cal E_\mu$ is a constant. $J_{kn}$ is then related to the probability
of the transition $|k>\to |n>$ induced by the perturbation $\delta H$,
($\omega \to 0$ at the end). This
suggests that in the fictitious 4+1 dimensional space-time, the dynamics of
SB$\chi$S might be related to transport properties of
massless quarks in a random medium characterized by a coloured magnetic
type disorder $G^a_\mu (x)$ with the probability distribution
(\ref{eq_ave}). A deeper insight into this formal analogy may be obtained
reexpressing the order parameters (\ref{eq_cond}) and (\ref {eq_dec}) in terms
of global characteristics of the infrared part of the Dirac spectrum.
Denoting by $\sum_{n}^\epsilon $
the sum over all states with ``energy ``  $|\lambda|<\epsilon $,
Eq. (\ref {eq_dec}) can be written as
\begin{equation}  \label{eq_twoep}
  F_0 ^2= {\pi}^2 \lim_{\epsilon,\epsilon^{\prime}\to 0} \lim_{L\to \infty}
\frac {1}{4\epsilon \epsilon^{\prime} L^4} <<\sum_{k}^\epsilon \sum_{n}^{
\epsilon^{\prime}} J_{kn} >>.  
\end{equation}
Setting $ \epsilon = \epsilon'$ in this equation, then dividing and
multiplying by
the square of the average number of states with $|\lambda | <\epsilon $,
$N(\epsilon ,L ) = << \sum_{n}^\epsilon >> $ , one finally obtains
\begin{equation} \label{eq_kubo}
F_0 ^2 ={\pi}^2 \lim_{\epsilon \to 0} \lim_{L \to \infty} L^4 J(\epsilon,L )
\rho^2(\epsilon ,L ).
\end{equation}
Here, $J(\epsilon,L)$ stands for the transition probability (\ref {eq_JKN})
averaged over all pairs of initial and final states with energy
$|\lambda|<\epsilon $ and over the disorder,
\begin{equation} \label{eq_mob}
J(\epsilon,L)= \frac {1}{N^2(\epsilon,L)} <<\sum_{nk}^{\epsilon} J_{nk} >>.
\end{equation}
$J$ measures the ``mobility'' of quarks at the infrared edge of the Dirac
spectrum, subject to the action of the electric field. $\rho(\epsilon,L)$
in Eq. (\ref {eq_kubo}) denotes the density of states, i.e. the number of
states per unit ``energy''  $\epsilon$ and unit volume: 
$\rho = N/2\epsilon L^4$.
 It is well known \cite {BanksCasher80} that the density of states defines
the condensate (\ref {eq_cond}):
\begin{equation} \label{eq_BC}
< \bar q q > = - \pi \lim_{\epsilon \to 0} \lim_{L \to \infty} \rho(\epsilon
,L) .
\end{equation}
Eq. (\ref {eq_kubo}) can be viewed as an ultrarelativistic ($m=0$) version
of the Kubo-Greenwood formula for electric conductivity (see e.g.
\cite {Mott70},\cite {Thouless74},\cite {XYZ} ). The latter usually deals
with the non-relativistic current $\phi ^{\dagger} \nabla _{\mu} \phi /m $
(instead of $\phi ^{\dagger} \gamma _{\mu} \phi $ in Eq (\ref {eq_JKN})) and
it involves the neighbourhood of the Fermi energy instead of the infrared end
of the Dirac spectrum. The relativistic transition probability $J_{kn}$
satisfies for each gauge field configuration a completeness sum rule
$\sum_ {n} J_{kn} = 1 $ implying that the ``mobility'' $J(\epsilon , L)$
should be bounded by the inverse number of states:
\begin{equation} \label{eq_bound}
J(\epsilon , L ) \le \frac{1}{N(\epsilon ,L)}.
\end{equation}
There is no analogous restriction in the non-relativistic theory of
conductivity.

We now turn to the double limit that appears in the expressions (\ref
{eq_kubo}) and (\ref{eq_BC}) of the order parameters $F_0 ^2$ and
$<\bar q q>$. SB$\chi$S
can only occur if the infinite volume limit is performed first. The
result then depends on the degree of accumulation of small eigenvalues
$\lambda_{n}$ around 0 as $L \to \infty$. The latter may be characterized
by an index $\kappa$ indicating how rapidly the n-th (positive) eigenvalue
averaged over the disorder vanishes, if $L \to \infty $ and n is kept fixed:
We shall say that $\lambda_{n}$ belongs to the ``$\kappa $-band'' if
$<<\lambda_{n}>>$ behaves for large $L$  as $L^{-\kappa}$. We assume that the
spectrum of the Dirac operator consists of several ``$\kappa$-bands''. The
case $\kappa =1$ is
characteristic of perturbation theory and it has been shown \cite{VW84}
that $\kappa \ge 1$. We are interested in small
eigenvalues that belong to the maximal $\kappa$-band. The corresponding
index $\kappa$
determines the leading behavior of the number of states $N(\epsilon,L)$
for large $L$ and small $\epsilon$: It follows from the definition that
the number of states in a $\kappa$-band becomes a function of a single
scaling variable $\epsilon L^{\kappa}$. Since on the other hand, one
expects the number of states $N(\epsilon,L)$ to be proportional to the
volume $L^4$, one should have
\begin{equation} \label{eq_ns}
N(\epsilon , L ) = (\frac {2 \epsilon}{\mu})^{4/\kappa} (\mu L)^4 +
\ldots ,
\end{equation}
 where $\mu$ is a mass scale and the dots stand for higher powers of
 $\epsilon$ .This formula has three important consequences: i) The first one
concerns the condition of existence  of SB$\chi$S , i.e. the requirement
that the limit in Eq (\ref{eq_kubo}) be non-vanishing and finite. Using
inside Eq. (\ref {eq_kubo}) the unitarity bound (\ref {eq_bound}), one obtains
\begin{equation} \label{eq_kappa2}
F_0 ^2 \le \pi ^2 \mu ^2 \lim_{\epsilon \to 0} (\frac {2 \epsilon}{\mu})^{
4/\kappa  - 2}.
\end{equation}
Hence, SB$\chi$S can take place only provided  $\kappa \ge 2 $. ii) The
second consequence concerns the existence of a non-zero quark condensate. Using
Eq. (\ref {eq_ns}) in the formula (\ref {eq_BC}), one gets
\begin{equation} \label{eq_kappa4}
<\bar q q > = - \pi \mu ^3 \lim_{\epsilon \to 0} (\frac {2 \epsilon}{\mu})^{
4/\kappa  - 1}  .
\end{equation}
Hence, the chiral condensate is formed if and only if $\kappa =4 $, a known
result obtained in \cite {LS}. The case $\kappa > 4$ would lead to an infinite
condensate and it can likely be excluded \cite {Gasser97}. Indeed, the well
known Ward identity leads to the expression
\begin{equation} \label{eq_ward}
<\bar q q> \delta^{ij} = \lim_{m \to 0} k_\mu \int dx e^{ikx} <A^i _\mu (x)
P^j (0)>
\end{equation}
and one usually argues that the zero mass limit of the two point function
on the r.h.s. of this equation should exist at least for some (non-exceptional)
momentum transfer $k_\mu$. Hence, $\kappa > 4$ would represent a too strong
infrared singularity incompatible with general properties of a field theory.
iii) The third consequence of Eq (\ref{eq_ns}) concerns the properties of
the effective theory which describes the low energy dynamics in terms of
Goldstone boson fields, whenever SB$\chi$S occurs, i.e. for $\kappa \ge 2$.
One usually assumes that the corresponding effective Lagrangian
\cite{Weinberg79}, \cite{GL84} , is analytic in Goldstone
boson fields and in scalar sources, i.e. in the quark mass. In this case, the
leading small $m$ behaviour of the partition function (given by the tree
approximation) is necessarily analytic and the same should be true for
the leading small $\epsilon$ behavior of the number of states $N(\epsilon ,L )$
. This requires $4/{\kappa}$=integer, selecting the values
$\kappa = 4$ and $\kappa = 2$ as the two candidates leading to SB$\chi$S
and to  an analytic effective Lagrangian. For  $\kappa=2 $, the
density of states is suppressed, $\rho(\epsilon, L) = 2 \epsilon {\mu}^2 $
and, consequently, the condensate $<\bar q q>$ vanishes.
On the other hand,  $\kappa=4$ is characterized by
a constant $\rho$ leading to a non-vanishing condensate
$<\bar q q> = - \pi  {\mu}^3$. A given $\kappa$-band ($\kappa$ = 2 or 4)
contributes to $F_0 ^2$ if the corresponding density of states is
modulated by an appropriate quark mobility $J(\epsilon,L)$ following
Eq. (\ref{eq_kubo}):  For $\kappa =2$, SB$\chi$S can take place despite
$<\bar q q>=0 $, provided the small density of states is compensated by a
high mobility, behaving typically like the inverse number of states.
In the $\kappa =4$ - band, the mobility should on the other hand be
suppressed by a factor of volume and should be independent of $\epsilon$.
One can, at least, conceive a class of models which realize the
different infrared behavior of $J(\epsilon ,L)$ as required by the two
alternatives of SB$\chi$S. The guideline will be a possible connection
between mobility and localization properties \cite{Anderson} of Euclidean
low energy Dirac eigenstates $\phi_{n}(x)$ suggested by the
non-relativistic theory of transport in disordered systems (see
\cite{xyz} for a recent review).

{\bf 1) The $\kappa =2$ - band}. Since the density of states is suppressed,
it is more convenient to use a particular form of Eq. (\ref{eq_kubo})
\begin{equation}  \label{eq_prob}
F_0 ^2 = {\pi}^2 {\mu}^2 \lim_{\epsilon \to 0} \lim_{L\to \infty}
N(\epsilon,L) J(\epsilon,L)
\end{equation}
which holds for $\kappa =2$. Notice that $\sigma (\epsilon ,L) = N J$
represents the total probability of transition from a given initial state
$|i>$ with $|\lambda|<\epsilon$ to any state with $|\lambda|<\epsilon$.
The upper bound (\ref{eq_bound}) implies $\sigma \le 1 $. Now, let us assume
that the transition between any two states from the $\kappa =2$ - band
can occur with a non-negligible intensity $J_{kn}$ , provided the two states
are close enough in energy. This is conceivable for {\bf extended
(delocalized) states}. Furthermore, transitions between the infrared edge
of the $\kappa =2$ - band and the perturbative $\kappa =1$ - band (which is
always present) should be suppressed because of a large difference in energy.
Under these circumstances, one may expect that for $|\lambda_i| <\epsilon$
the sum $\sum_{f}J_{if}$ will be dominated by states with energy
$|\lambda|<\epsilon $, whereas the contribution of remaining states will be
relatively suppressed as $N(\epsilon,L) \to \infty $. In this case, the bound
$\sigma(\epsilon, L) \le 1 $ should be nearly saturated : $ \sigma (\epsilon ,
L) = 1 - 0(1/N) $. Only a much weaker form of this assumption is actually
needed: It is sufficient that for  large L and  small $\epsilon$,
$\sigma(\epsilon ,L))$ admits an expansion
\begin{equation}  \label{eq_exp}
\sigma (\epsilon , L) = N J = w(1/N) + 2 \tilde w (1/N) \epsilon + \cdots
\end{equation}
with $0 < w(0) \le 1 $. (The particular case described above corresponds to
$w(0)= 1 $ .) Notice that this expansion, as well as previous reasoning ,
only applies if the number of states $N(\epsilon ,L)$ is sufficiently large.
In particular, Eq. (\ref{eq_exp}) contains no information about the limit
$\epsilon \to 0 $ , $L$ fixed , in which $\sigma$ should vanish (chiral
symmetry should be restored) since there are no states with
$|\lambda|<\epsilon $. The final result then reads
$F_0 ^2 = {\pi}^2 {\mu}^2 w(0)$,
where $w(0)$ is the probability defined in Eq. (\ref{eq_exp}) and $\mu$ is
the mass scale that controls the leading dependence of the number of states
N on $\epsilon$ and on $L$. Assuming the latter to be of the order of the
intrinsic QCD scale $\Lambda_{QCD} \sim 300$MeV , one obtains $w(0) \sim
0.01 $ , whereas $w(0) \le 1$ corresponds to $\mu \ge 30$ MeV .

{\bf 2) The $\kappa =4$ - band.} Since the density of states stays constant,
Eq (\ref{eq_kubo}) requires $L^4 J(\epsilon, L)$ to be independent of
$\epsilon$ and of $L$ . This suppression of mobility by the factor of volume
hardly fits into the previous picture, in which all states with
$|\lambda | < \epsilon $ participated in the transition. Instead, the
intensity $J_{kn}$ should now be suppressed for most pairs of initial and
final states, even if they are close in energy. The following scenario
attempts to explain naturally such a suppression as a consequence of a kind
of {\bf localization of states} in the $\kappa =4 $ - band.

i)Let us first assume that to each state $|n>$ , one can associate 
a ``center `` $C_n$
localized at a point $x_n$, all centers forming a (random) lattice
$ Z_x ^4 $ covering the hypercube $L \times L \times L \times L $. The
centers $C_n$ may be viewed as localized defects or instantons \cite{Dyako}
their precise nature and origin are at this stage irrelevant.

ii) Next, we assume that there is a one-to-one correspondence between states
from the infrared edge of the $\kappa =4 $ - band and the points of the
five-dimensional lattice $ Z_{\lambda} \times Z_x ^4 $, $Z_{\lambda}$
collecting the levels $\lambda_{n} $ . Notice that i) and ii) yield the
correct number of states : $N(\epsilon, L) = 2 \epsilon {\mu}^3 L^4 $.

iii)Finally, $J_{kn}$ will be assumed to be independent of energy and
sufficiently suppressed if the distance $ r_{kn} =|\vec x_k - \vec x_n | $
between the corresponding two centers $ C_k$ and $C_n$ grows. Only a rather
weak form of this suppression is needed: Writing, for example,
$J_{kn} = F(r_{kn})$ , the function $F$ should be integrable in the infinite
volume limit , $ lim_{L \to \infty} \int d^4 x F(r) = l^4 $. The new
length scale $l$ characterizes the localization of the low $\lambda$ states
from the $\kappa =4$ - band. It is now straightforward to use
Eq. (\ref{eq_mob}) and to estimate the mobility
\begin{equation} \label{eq_loc}
J(\epsilon, L) = 2 \epsilon {\mu}^3 \frac {l^4}{N} = \frac {l^4}{L^4},
\end{equation}
making apparent the required suppression by the factor of volume. The same
result can be obtained from the expansion (\ref{eq_exp}) setting
$w(1/N)=0$ and writing $\tilde w (0) = {\mu}^3 l^4 $ . Hence, the above
argument of localization provides a justification of the absence of the first
term in the expansion (\ref{eq_exp}) in the case $\kappa =4$ . The Kubo-
type formula (\ref{eq_kubo}) now reduces to the simple relation
\begin{equation} \label{eq_fin}
F_0 = \pm  <\bar q q >   l^2,
\end{equation}
where $l$ is the characteristic length of localization defined above. 
If the latter happens
to diverge ($ l \sim L $), states become delocalized , the condensate must
vanish and one recovers the $\kappa =2 $ case discussed before.

It is worth noting that the alternative of vanishing condensate $<\bar q q>$
(i.e. $\kappa=2$) does not imply the vanishing of all local condensates
of the type $\bar q q$. This is best illustrated by the case of the dimension
5 ``mixed condensate `` which can be expressed
by a formula similar to Eq. (\ref{eq_kubo})
\begin{equation} \label{eq_mixed}
<\bar q \sigma_{\mu\nu}G_{\mu\nu} q> = - \pi \lim_{\epsilon \to 0} 
\lim_{L \to \infty} G_{\Vert}(\epsilon,L) \rho (\epsilon,L),
\end{equation}
where $G_{\Vert}$ denotes the mean value of $\sigma_{\mu\nu}G_{\mu\nu}(x)$
in a state $\phi_{n}(x)$ with $|\lambda_{n}|<\epsilon$, averaged over all such
states and over the disorder. For $\kappa=2$, the mixed condensate will
remain non-zero, provided $G_{\Vert}(\epsilon,L)$ blows up as $1/\epsilon$,
indicating a possible importance of a large gauge field configurations
parallel to the quark
spin at the infrared edge of the $\kappa=2$ band. (In the
case $\kappa=4$, a constant $G_{\Vert}$ is sufficient to garantee a finite
mixed condensate.) Let us recall that a non-vanishing of some local
condensates of the type (\ref{eq_mixed}) is required by the consistency of
the pattern of SB$\chi$S with the short distance properties of the theory in
the limit $N_c \to \infty$  \cite{KdR} .

In QCD, the theoretical question of the importance of $<\bar q q>$ is of a
direct phenomenological relevance. The alternative of vanishing (or tiny)
condensate is at present not ruled out by available experimental data (see
\cite{JS} and references therein). The decisive test of the role of $<\bar q
q>$in SB$\chi$S \cite{pipi} will hopefully come from new high precision low
energy $\pi \pi$ scattering experiments \cite{Stern91}, currently under
preparation. We believe that the existence of theoretical alternatives
described
in this Letter demonstrates the fundamental importance of these and similar
experimental tests.

Stimulating discussions with A.Comtet, J.Gasser, M.Knecht, H.Leutwyler,
B.Moussallam, A.Smilga, D.Vautherin and J.Watson have been profitable.

\end{document}